\documentclass[11pt]{article}
\usepackage{float}
\usepackage{eurosym}
\usepackage{graphicx}
\usepackage{amsmath}
\usepackage{amssymb}
\usepackage{latexsym}
\usepackage{color}
\usepackage{epstopdf}
\usepackage[utf8]{inputenc}
\usepackage[T1]{fontenc}

\providecommand{\openone}{\leavevmode\hbox{\small1\kern-4.3pt\normalsize1}}

\begin{document}

\thispagestyle{empty}

\begin{center}
\vspace{1.8cm}

{\Large \textbf{Quantum Fisher information matrix in Heisenberg XY model}}

\vspace{1.5cm}

\textbf{L. Bakmou}$^{a}$ {\footnote{%
email: \textsf{baqmou@gmail.com }}}, \textbf{A. Slaoui}$^{a}$ {\footnote{%
email: \textsf{abdallahsalaoui1992@gmail.com; abdallah.slaoui@um5s.net.ma}}}, \textbf{M. Daoud}$^{b}$ {\footnote{%
email: \textsf{m\_daoud@hotmail.com}}} and \textbf{R. Ahl Laamara}$^{a, c}$ {%
\footnote{%
email: \textsf{ahllaamara@gmail.com}}}

\vspace{0.5cm}

$^{a}$\textit{LPHE-MS, Faculty of Sciences, Mohammed V University of Rabat, Rabat, Morocco}\\[0.5em]
$^{b}$\textit{Department of Physics, Faculty of Sciences, University Ibn Tofail, Kenitra, Morocco}\\[0.5em]
$^{c}$\textit{Centre of Physics and Mathematics (CPM), Mohammed V University
of Rabat, Rabat, Morocco}\\[0.5em]

\end{center}
\baselineskip=18pt
\medskip
\vspace{1cm}

\begin{abstract}

The quantum Fisher information matrix provides us with a tool to determine the precision, in any multiparametric estimation protocol, through quantum Cramér–Rao bound. In this work, we study simultaneous and individual estimation strategies using the density matrix vectorization method. Two special Heisenberg $XY$ models are considered. The first one concerns the anisotropic $XY$ model in which the temperature $T$ and the anisotropic parameter $\gamma$ are estimated. The second situation concerns the isotropic $XY$ model submitted to an external magnetic field $B$ in which the temperature and the magnetic field are estimated. Our results show that the simultaneous strategy of multiple parameters is always advantageous and can provide a better precision than the individual strategy in the multiparameter estimation procedures.\\

\end{abstract}

\newpage

\section{Introduction}
The parameter estimation is of paramount importance in the development of high precision devices in several areas of technology \cite{Giovannetti2006,Giovannetti2011,Helstrom1976}. Recently, quantum metrology has attracted considerable attention by employing the quantum effects to improve the precision limit and to develop new methods to measure physical parameters beyond the classical metrological methods \cite{Huelga1997,Escher2011}. Now, there exists several applications of quantum metrology. One may quote clock synchronization \cite{Joza2000}, the maximization of the sensitivity of gravitational waves detector \cite{Abbott2016}, the obtention of the bounds on the optimal estimation of phases \cite{Ballester2004,Monras2006,Aspachs2009}, the estimation of space–time parameters \cite{Nation2009,Weinfurtner2011,Aspachs2010}, electromagnetic field sensing \cite{Wasilewski2010,Cai2013} and the optimal estimation of the reservoirs temperature \cite{Monras2011,Correa2015,Boss2017}. The quantum metrology protocols can substantially improve the estimation precision by taking advantage of quantum correlations existing in a multipartite system: entanglement \cite{Einstein1935,Bell1964,Wootters1997}, quantum discord \cite{Giorda2010,Ollivier2001,Mancino2018}. \par
The limit of the precision  of measurement of a set of parameters ${\hat \theta }$ in quantum metrology is usually framed by the inequality called the quantum Cramér–Rao bound \cite{Paris2009} which writes ${\rm Cov}\left( {\hat \theta } \right) \ge {F^{ - 1}}$, where ${\rm Cov}\left( {\hat \theta } \right)$ denotes the covariance matrix of an estimation vector which contains the parameters to be estimated and $F$ denotes the quantum  Fisher information matrix (QFIM) \cite{Braunstein1994}. Obviously the Cramer–Rao inequality reduces, in the case of a single parameter $\theta$, to ${\rm Var}\left( \theta  \right) \ge {\mathcal{F}^{ - 1}}$,  where ${\rm Var}\left( \theta  \right)$ is the variance that corresponds to the square of the standard deviation and $\mathcal{F}$ denotes the quantum Fisher information quantity (QFI) \cite{Kay1993,Helstrom1976,Holevo2001}. More precision is obtained for small variance. So, the ultimate goal in any quantum metrology protocol is to reach the smallest value of the variance. In this direction, it is clear that the inverse of QFIM for many unknown parameters (or the inverse of the quantity QFI for one unknown parameter) provides the lower error limit of the estimation the parameters. Therefore, the way to increase QFIM (or QFI) is a challenging issue in quantum metrology \cite{Paris2009}. The Cramer–Rao inequality associated with the estimation of a single parameter has been studied extensively \cite{Genoni2013,Humphreys2013}. The single-parameter estimation plays an important role in many ways, because of the existence of an optimal probe state containing a maximum amount of QFI \cite{Yuan2017,Liu2013,Yuen1973}. Realistic problems can usually involve several parameters, because there is no optimal probe state in which QFIM is larger than the other states \cite{Matsumotoar2002}. In addition, the Cramer–Rao inequality is not always saturable because the measurements for different parameters may be incompatible \cite{Rehacek2018,Ragy2016}. Therefore, the simultaneous estimation of several parameters becomes an important task in quantum metrology. Recently, studies on multiparameter estimation have attracted a great deal of interest. Simultaneous estimation of several parameters can give a better precision than their individual estimation. It has been shown that entanglement of several particles can improve multiphase estimation processes \cite{Spagnolo2012,Zhang2017}. In addition, two-mode entangled coherent states are proposed for estimating linear and nonlinear phase shifts \cite{Cheng2014}. \par
It has been recently reported in several studies that quantum correlations present in a physical system can be considered as an essential resource to improve precision in quantum metrology \cite{Pezze2009}. In this respect the investigation whether the increase in QFIM (or QFI) can be used as indicator of the existence of quantum correlations in a multipartite system and to quantify the quantum correlations in terms of quantum Fisher information. Some works were developed in this direction \cite{Rivas2010,Li2013,Girolami2014}. However, it must be stressed the understanding of quantum correlations and entanglement in quantum metrology and their role in achieving the highest precision of the estimated parameter, is far from complete. \par
This paper is structured as follows. In Section 2, we discuss the basics of multiparameter quantum estimation theory and the essential mathematical tools to derive the quantum Fisher information matrix. A special attention is devoted to the vectorization method of density matrix. In Section 3, we examine the precision of the multiparametric estimation using QFIM in Heisenberg $XY$ model. We consider two special situations \cite{Zhang2007}. The first one concerns the anisotropic $XY$ model and the second deals with the isotropic $XY$ model submitted to an external magnetic field. We derive the corresponding symmetric logarithmic derivatives and conditions for saturability of the quantum Cramér–Rao bound which gives the ultimate precision. We also analyze the simultaneous and individual strategies. This is done by introducing the ratio between the minimal amounts of total variances for each estimating protocol. We end up the paper with concluding remarks.
\section{Quantum Fisher information matrix}
In this section, we review some mathematical tools that are essential to derive the quantum Fisher information matrix. In this sense we consider an algebraic application that transforms a matrix into a column vector to define the elements of the quantum Fisher information matrix without diagonalizing the density matrix. Let ${\mathbb{M}^{n \times n}}$ denote the space of $n \times n$ real (or complex) matrices. For any matrix $A \in {\mathbb{M}^{n \times n}}$, the vec-operator is defined as \cite{Gilchrist}
\begin{equation}
{\rm vec}\left[ A \right] = {\left( {{a_{11}},...,{a_{n1}},{a_{12}},...,{a_{n2}},...,{a_{1n}},...,{a_{nn}}} \right)^T}.
\end{equation}
Furthermore, using the expression $A = \sum\limits_{k,l = 1}^n {{a_{kl}}} \left| k \right\rangle \left\langle l \right|$, the vec-operator rewrites
\begin{equation}
{\rm vec}\left[ A \right] = \left( {{\mathbb{I}_{n \times n}} \otimes A} \right)\sum\limits_{i=1}^n {{e_i} \otimes {e_i}}, \label{vecA}
\end{equation}
where ${e_i}$ denotes the elements of the computational basis of ${\mathbb{M}^{n \times n}}$. This means that the vec-operator creates a column vector from a matrix $A$ by stacking the column vectors of $A$ below one another. Using the properties of the Kronecker product \cite{Schacke2004}, one gets
\begin{equation}
{\rm vec}\left[ {AB} \right] = \left( {{\mathbb{I}_n} \otimes A} \right){\rm vec}\left[ B \right] = \left( {{B^T} \otimes {\mathbb{I}_n}} \right){\rm vec}\left[ A \right], \label{vec1}
\end{equation}
\begin{equation}
{\rm tr}\left( {{A^\dag }B} \right) = {\rm vec}{\left[ A \right]^\dag }{\rm vec}\left[ B \right]. \label{vec3}
\end{equation}
\begin{equation}
{\rm vec}\left[ {AXB} \right] = \left( {{B^T} \otimes A} \right){\rm vec}\left[ X \right],\label{vec2}
\end{equation}
for any matrices $A$, $B$ and $X$. Before giving the explicit expression of the quantum Fisher information matrix using the vec-operator associated to the density matrix $\rho$, we will review some elements of the concept of quantum Fisher information matrix (QFIM) to estimate several parameters $\left\{{\theta _i}\right\} = \left\{ {{\theta _1},{\theta _2},...,{\theta _n}} \right\}$. The quantum Fisher information is the maximum amount of information about estimating a parameter obtained from optimal measurements. For states ${\rho_\theta }$, dependent on a single parameter $\theta$, the quantum Fisher information is defined by $\mathcal{F}\left( {{\rho _\theta }} \right) = {\rm Tr}\left\{ {{\rho _\theta }{L_\theta }^2} \right\}$, where $L_\theta$ is the symmetric logarithmic derivative. In situations where more parameters ${\theta _i}$ are involved, the relevant object, in the estimation problem, is given by the so-called quantum Fisher information matrix \cite{Paris2009}
\begin{equation}
{F_{ij}} = \frac{1}{2}{\rm Tr}\left\{ {\left( {{{\hat L}_{\theta _i}}{{\hat L}_{\theta _j}} + {{\hat L}_{\theta _j}}{{\hat L}_{\theta _i}}} \right) \rho } \right\}, \label{F}
\end{equation}
where the symmetric logarithmic derivatives ${{{\hat L}_{\theta _i}}}$ satisfy the algebraic equations:
\begin{equation}
2{\partial _{\theta _i}} \rho  = {\hat L_{\theta _i}} \rho  + \hat \rho {\hat L_{\theta _i}}. \label{L}
\end{equation}
Clearly, the explicit derivation of the quantum Fisher information matrix (\ref{F}) requires the expression of the symmetric logarithmic derivative ${{{\hat L}_{\theta _i}}}$ (\ref{L}). The
explicit expressions of the quantum Fisher information matrix have been reported in the literature \cite{Banchi2014,Sommers2003,Paris2009}. Using the spectral
decomposition of the density matrix, i.e, ${\rho } = \sum\limits_k
{{p_k}} \left| k \right\rangle \left\langle k \right|$, the quantum Fisher information matrix was derived in terms of the eigenvalues of ${\rho }$  \cite{Banchi2014,Sommers2003}
\begin{equation}
    {F_{ij}} = 2\sum\limits_{{p_k} + {p_l} > 0} {\frac{{\left\langle k \right|{\partial _{{\theta _i}}}{\rho }\left| l \right\rangle \left\langle l \right|{\partial _{{\theta _j}}}{\rho }\left| k \right\rangle }}{{{p_k} + {p_l}}}}, \label{F1}
\end{equation}
and the symmetric logarithmic derivatives are given by
\begin{equation}
    {L_{{\theta _i}}} = 2\sum\limits_{{p_k} + {p_l} > 0} {\frac{{\left\langle k \right|{\partial _{{\theta _i}}}{\rho }\left| l \right\rangle }}{{{p_k} + {p_l}}}} \left| k \right\rangle \left\langle l \right|.
\end{equation}
The quantum Fisher information matrix can be written in terms of the exponentiation of the density matrix as \cite{Paris2009}
\begin{equation}
    {F_{ij}} = 2\int\limits_0^\infty  {{\rm Tr}\left[ {{e^{ - {\rho }t}}{\partial _{{\theta _i}}}{\rho }{e^{ - {\rho }t}}{\partial _{{\theta _j}}}{\rho }} \right]}. \label{F2}
\end{equation}
Very recently, a new explicit expression of the quantum Fisher information matrix, based on the vectorization method of density matrix ${\rho}$, has been introduced in \cite{Safranek2018}. This method has the advantage of being analytically computable for an arbitrary system. It does not require the diagonalization the density matrix as in the expression (\ref{F1}) or to compute the integral and exponentiation as in the equation (\ref{F2}). It is based on the computation of the inverse of the following matrix
\begin{equation}
\Lambda  = \left( {{\rho ^T} \otimes \mathbb{I} + \mathbb{I} \otimes \rho } \right). \label{landa}
\end{equation}
Using the properties given by the equations (\ref{vec1}), (\ref{vec2}) and (\ref{vec3}), it is easy to check that the quantum Fisher information matrix, given by the equations (\ref{F1}) and (\ref{F2}), rewrites as
\begin{equation}
{F_{ij}} = 2{\rm vec}{\left[ {{\partial _i}\hat \rho } \right]^T}{\Lambda ^{ - 1}}{\rm vec}\left[ {{\partial _j}\hat \rho } \right]. \label{Flanda}
\end{equation}
and the symmetric logarithmic derivatives are given by
\begin{equation}
{\rm vec}\left[ {L_{{\theta _i}}} \right] = 2{\Lambda ^{ - 1}}{\rm vec}\left[ {{\partial _i}\hat \rho } \right]. \label{vecL}
\end{equation}
Usually, in single parameter estimation scenarios, the scalar Cramer-Rao inequality ${\rm Var}\left( \theta  \right) \ge {\mathcal{F}^{ - 1}}$  is always saturable. This saturation gives an optimal quantum measurement operators which is obtained by the projectors on the eigenvectors of the symmetric logarithmic derivative operators ${L_\theta }$.  Unlike the single parameter estimation, the matrix Cramer-Rao inequality  in  multiparameter estimation scenarios,  ${\rm Cov}\left( {\hat \theta } \right) \ge {F^{ - 1}}$, can not always be saturable. This is due to the fact that the optimal operators measurements of different parameters can be incompatibles \cite{Rehacek2018,Ragy2016}. Therefore, it is natural look for the conditions that must be verified such that this inequality can be saturated. For this end, Eq.(\ref{vecL}) must be solved to determine the symmetric logarithmic derivatives $L_{{\theta _i}}$ corresponding to the different estimated parameters. In the case where the operators $L_{{\theta _i}}$ commute, one can find a common eigenbasis for all symmetric logarithmic derivatives. In this picture one can perform simultaneous measurement saturating the Cramer-Rao inequality. The commutativity condition $\left[ {L_{{\theta _i}},L_{{\theta _j}}} \right] = 0$ is sufficient but not necessary. In the case where the symmetric logarithmic derivatives are not commuting, the condition $Tr\left( {\rho \left[ {L_{{\theta _i}},L_{{\theta _j}}} \right]} \right) = 0$ ensures  the saturation of Cramér–Rao inequality \cite{Ragy2016,Matsumoto2002,Crowley2014}.
\section{QFIM in the quantum Heisenberg XY model}
The Heisenberg Hamiltonian for a chain of $N$ qubits, interacting with the nearest neighbor, can be written as \cite{Korepin1993,Wang2002,Kamta2002}
\begin{equation}
H = \sum\limits_{n = 1}^N {\left( {{J_x}S_n^xS_{n + 1}^x + {J_y}S_n^yS_{n + 1}^y + {J_z}S_n^zS_{n + 1}^z} \right)} \label{HN}
\end{equation}
where $S_n^i = \frac{1}{2} \sigma _n^i\,$( $i = x,y,z$ ) and $\sigma _n^i$ are the local spin-$\frac{1}{2}$ operators and Pauli  matrices respectively at site $n$. We assume periodic boundary conditions, i.e., $S_{N + 1}^i = S_1^i$. The parameters ${J_i}$ denote the coupling constants for the spin interaction. For ${J_x} \ne {J_y} \ne {J_z}$, the Heisenberg chain is called $XYZ$ model. In the particular cases ${J_x} = {J_y} \ne {J_z}$ and ${J_x} = {J_y} = {J_z}$ are the $XXZ$ and $XXX$ models respectively. The chain is antiferromagnetic (AFM) for ${J_i}$ positive, and ferromagnetic (FM) for ${J_i}$  negative (see \cite{Ha1996} and references therein). The Heisenberg antiferromagnetic and ferromagnetic properties have been considered in the context of quantum information science \cite{DiVincenzo2000,Loss1998}. The interest for this system has been  revived thanks to several proposals for the realization of solid state quantum computation processors using electron spin localized as qubits. In such realizations the basic gate operations involve different forms of the  Heisenberg Hamiltonian. In this context the $XY$ (${J_z} = 0$) and Ising (${J_y} = {J_z} = 0$) interactions are analyzed in the references \cite{Imamoglu1999,Raussendorf2001}. The role of QFI to detect the quantum phase transition (QPT) was investigated recently in the literature. For instance, in \cite{Ye2016}, the authors have analyzed the phase transition in $XY$ spin models. In particular, it has been shown that the first and second derivatives of QFI versus the QPT parameter (the magnetic field), in the thermodynamic limit (i.e., $N\longrightarrow\infty$), exhibit a local minimum near the critical point. Furthermore, the first derivative displays the phenomenon of sudden transition and the second derivative represents the sudden jump and divergence at the critical point transition. In this work, we will use the quantum Fisher information matrix (QFIM) to study the precision of the measurement of some parameter occurring in the $XY$ anisotropic model and the $XY$ isotropic model in a external magnetic field ( along the z axis). The state of a quantum system described by the Hamiltonian $H$ at thermal equilibrium is described by the Gibb's density operator, $\rho  = {{{\exp\left( - \beta H\right)}} \mathord{\left/
        {\vphantom {{{e^{ - \beta H}}} Z}} \right.
        \kern-\nulldelimiterspace} Z}$ where $Z = {\rm Tr }{e^{ - \beta H}}$ is the partition function  of the system and $\beta  = {1 \mathord{\left/
{\vphantom {1 {kT}}} \right.
\kern-\nulldelimiterspace} {kT}}$, $k$ is Boltzmann constant which we henceforth will take to 1 and $T$ the temperature.
\subsection{Anisotropic XY model}
We first consider the Hamiltonian $H$ for two-qubit Heisenberg $XY$ model (i.e., $N=2$ and $J_z=0$). Using the raising and lowering operators  $\sigma _n^ \pm  = \sigma_n^x \pm \sigma_n^y$, the Hamiltonian (\ref{HN}) gives
\begin{equation}
H = J\left( {\sigma _1^ + \sigma _2^ -  + \sigma _2^ + \sigma _1^ - } \right) + J\gamma \left( {\sigma _1^ + \sigma _2^ +  + \sigma _2^ - \sigma _1^ - } \right),
\end{equation}
where $J = \frac{{{J_x} + {J_y}}}{2}$ and $\gamma  = \frac{{{J_x} - {J_y}}}{{J_x} + {J_y}}$. Without loss of generality, we set $J=1$. The parameter $\gamma$ is the anisotropy parameter. It is zero (${{J_x} = {J_y}}$) for the isotropic $XX$ model and $ \pm 1$ for the Ising model. The eigenvalues and eigenstates of the Hamiltonian $H$ are analytically obtained as  $H\left| {{\psi ^ \pm }} \right\rangle  =  \pm \left| {{\psi ^ \pm }} \right\rangle $, $H\left| {{\chi ^ \pm }} \right\rangle  =  \pm\gamma \left| {{\chi ^ \pm }} \right\rangle$, with the $\left| {{\psi ^ \pm }} \right\rangle  = \frac{1}{{\sqrt 2 }}\left( {\left| {01} \right\rangle  \pm \left| {10} \right\rangle } \right)$ and $\left| {{\chi ^ \pm }} \right\rangle  = \frac{1}{{\sqrt 2 }}\left( {\left| {00} \right\rangle  \pm \left| {11} \right\rangle } \right)$  are the maximally entangled Bell states. In the standard basis $\left\{ {\left| {00} \right\rangle ,\left| {01} \right\rangle ,\left| {10} \right\rangle ,\left| {11} \right\rangle } \right\}$, the density matrix $\rho  = {{{\exp\left( - \beta H\right)}} \mathord{\left/
        {\vphantom {{{e^{ - \beta H}}} Z}} \right.
        \kern-\nulldelimiterspace} Z}$ can be written as
\begin{equation}
    \rho  = \left[ {\begin{array}{*{20}{c}}
            a&0&0&x\\
            0&b&z&0\\
            0&z&b&0\\
            x&0&0&a
        \end{array}} \right], \label{gho1case}
    \end{equation}
    where the entries given by
    \begin{equation}
a = \frac{{\cosh\left( \beta\gamma \right)}}{{2\cosh\left( \beta\gamma \right) + 2\cosh\left( \beta \right)}}, \hspace{1.5cm}b = \frac{{\cosh\left( \beta \right)}}{{2\cosh\left( \beta\gamma \right) + 2\cosh\left( \beta \right)}},
    \end{equation}
    \begin{equation}
        x = \frac{{ - \sinh\left( \beta\gamma \right)}}{{2\cosh\left( \beta\gamma \right) + 2\cosh\left( \beta \right)}}, \hspace{1.5cm}z = \frac{{ - \sinh\left( \beta \right)}}{{2\cosh\left( \beta\gamma \right) + 2\cosh\left( \beta \right)}}.
    \end{equation}
In what follows, we shall consider the estimation of the anisotropy parameter $\gamma$ and the temperature $T$. To evaluate the quantum Fisher information matrix, one has first to compute the matrix $\Lambda$ given by (\ref{landa}). It writes

\begin{equation}
\Lambda  = \left[ {\begin{array}{*{20}{c}}
        {{\Lambda _{11}}}&{{0_{4 \times 4}}}&{{0_{4 \times 4}}}&{{\Lambda _{14}}}\\
        {{0_{4 \times 4}}}&{{\Lambda _{22}}}&{{\Lambda _{23}}}&{{0_{4 \times 4}}}\\
        {{0_{4 \times 4}}}&{{\Lambda _{32}}}&{{\Lambda _{33}}}&{{0_{4 \times 4}}}\\
        {{\Lambda _{41}}}&{{0_{4 \times 4}}}&{{0_{4 \times 4}}}&{{\Lambda _{44}}}
        \end{array}} \right], \label{LL}
\end{equation}
with $\Lambda _{ij}$ $(i,j=1,2,3,4)$ are the $4\times4$ matrix given by
\begin{equation}
{\Lambda _{11}} = {\Lambda _{44}} = \left[ {\begin{array}{*{20}{c}}
    {2a}&0&0&x\\
    0&{a + b}&z&0\\
    0&z&{a + b}&0\\
    x&0&0&{2a}
    \end{array}} \right], \hspace{1cm}{\Lambda _{22}} = {\Lambda _{33}} = \left[ {\begin{array}{*{20}{c}}
{a + b}&0&0&x\\
0&{2b}&z&0\\
0&z&{2b}&0\\
x&0&0&{a + b}
\end{array}} \right],
\end{equation}
and
\begin{equation}
    {\Lambda _{23}} = {\Lambda _{32}} = \left[ {\begin{array}{*{20}{c}}
        z&0&0&0\\
        0&z&0&0\\
        0&0&z&0\\
        0&0&0&z
        \end{array}} \right], \hspace{3cm} {\Lambda _{41}} = {\Lambda _{14}} = \left[ {\begin{array}{*{20}{c}}
x&0&0&0\\
0&x&0&0\\
0&0&x&0\\
0&0&0&x
\end{array}} \right].
\end{equation}
The inverse of the matrix $\Lambda$ Eq.(\ref{LL}) is given by
\begin{equation}
{\Lambda ^{ - 1}} = \left[ {\begin{array}{*{20}{c}}
        {{{\left( {{\Lambda ^{ - 1}}} \right)}_{11}}}&{{0_{4 \times 4}}}&{{0_{4 \times 4}}}&{{{\left( {{\Lambda ^{ - 1}}} \right)}_{14}}}\\
        {{0_{4 \times 4}}}&{{{\left( {{\Lambda ^{ - 1}}} \right)}_{22}}}&{{{\left( {{\Lambda ^{ - 1}}} \right)}_{23}}}&{{0_{4 \times 4}}}\\
        {{0_{4 \times 4}}}&{{{\left( {{\Lambda ^{ - 1}}} \right)}_{32}}}&{{{\left( {{\Lambda ^{ - 1}}} \right)}_{33}}}&{{0_{4 \times 4}}}\\
        {{{\left( {{\Lambda ^{ - 1}}} \right)}_{41}}}&{{0_{4 \times 4}}}&{{0_{4 \times 4}}}&{{{\left( {{\Lambda ^{ - 1}}} \right)}_{44}}}
        \end{array}} \right],
\end{equation}
with
\begin{equation}
{\left( {{\Lambda ^{ - 1}}} \right)_{11}} = {\left( {{\Lambda ^{ - 1}}} \right)_{44}} = \left[ {\begin{array}{*{20}{c}}
    \alpha &0&0&\xi \\
    0&\delta &\lambda &0\\
    0&\lambda &\delta &0\\
    \xi &0&0&\alpha
    \end{array}} \right], \hspace{1cm}
{\left( {{\Lambda ^{ - 1}}} \right)_{22}} = {\left( {{\Lambda ^{ - 1}}} \right)_{33}} = \left[ {\begin{array}{*{20}{c}}
    \delta &0&0&\varepsilon \\
    0&\nu &\mu &0\\
    0&\mu &\nu &0\\
    \varepsilon &0&0&\delta
    \end{array}} \right],
\end{equation}
and
\begin{equation}
    {\left( {{\Lambda ^{ - 1}}} \right)_{23}} = {\left( {{\Lambda ^{ - 1}}} \right)_{32}} = \left[ {\begin{array}{*{20}{c}}
        \lambda &0&0&\eta \\
        0&\mu &\omega &0\\
        0&\omega &\mu &0\\
        \eta &0&0&\lambda
        \end{array}} \right], \hspace{1cm} {\left( {{\Lambda ^{ - 1}}} \right)_{41}} = {\left( {{\Lambda ^{ - 1}}} \right)_{14}} = \left[ {\begin{array}{*{20}{c}}
\xi &0&0&\tau \\
0&\varepsilon &\eta &0\\
0&\eta &\varepsilon &0\\
\tau &0&0&\xi
\end{array}} \right],
\end{equation}

where the elements $\alpha$, $\xi $, $\delta$, $\lambda$, $\tau$, $\varepsilon$, $\eta$, $\upsilon$, $\mu$ and $\omega$ are respectively given by
\begin{equation}
    \alpha  = \frac{1}{4}\left( {\cosh\left( \beta \right) + \cosh\left( \beta\gamma \right)} \right)\left( {3 + \cosh\left( 2\beta\gamma \right)} \right){\mathop{\rm sech }\nolimits} \left( \beta\gamma \right),
\end{equation}
\begin{equation}
    \xi  = \frac{1}{2}\left( {\cosh\left( \beta \right) + \cosh\left( \beta\gamma \right)} \right){\mathop{ \sinh}\nolimits} \left( \beta\gamma \right),
\end{equation}
\begin{equation}
    \delta  = 1 + \cosh\left( \beta \right)\cosh\left( \beta\gamma \right), \hspace{1.5cm}\lambda  = \sinh\left( \beta \right)\cosh\left( \beta\gamma \right),
\end{equation}
\begin{equation}
    \tau  = \frac{1}{2}\left( {\cosh\left( \beta\gamma \right) + \cosh\left( \beta \right)} \right)\sinh\left( \beta\gamma \right)\tanh\left( \beta\gamma \right),
\end{equation}
\begin{equation}
    \varepsilon  = \cosh\left( \beta \right)\sinh\left( \beta\gamma \right), \hspace{1.5cm} \eta  = \sinh\left( \beta \right)\sinh\left( \beta\gamma \right),
\end{equation}
\begin{equation}
    \upsilon  = \frac{1}{4}\left( {3 + \cosh\left( 2\beta \right)} \right)\left( {\cosh\left( \beta \right) + \cosh\left( \beta\gamma \right)} \right){\mathop{\rm sech}\nolimits} \left( \beta \right),
\end{equation}
\begin{equation}
    \mu  = \frac{1}{2}\left( {\cosh\left( \beta \right) + \cosh\left( \beta\gamma \right)} \right){\mathop{\sinh}\nolimits} \left( \beta \right),
\end{equation}
\begin{equation}
    \omega  = \frac{1}{2}\left( {\cosh\left( \beta \right) + \cosh\left( \beta\gamma \right)} \right){\mathop{\rm \sinh}\nolimits} \left( \beta \right)\tanh\left( \beta \right).
\end{equation}
Using the definition (\ref{vecA}), one writes
\begin{equation}
{\rm vec}\left[ {{\partial _\gamma }\rho } \right] = {\left( {{\partial _\gamma }a,0,0,{\partial _\gamma }x,0,{\partial _\gamma }b,{\partial _\gamma }z,0,0,{\partial _\gamma }z,{\partial _\gamma }b,0,{\partial _\gamma }x,0,0,{\partial _\gamma }a} \right)^T},
\end{equation}
and
\begin{equation}
{\rm vec}\left[ {{\partial _T}\rho } \right] = {\left( {{\partial _T}a,0,0,{\partial _T}x,0,{\partial _T}b,{\partial _T}z,0,0,{\partial _T}z,{\partial _T}b,0,{\partial _T}x,0,0,{\partial _T}a} \right)^T}.
\end{equation}
Using (\ref{Flanda}), the quantum Fisher information matrix  can be determined as
\begin{equation}
    F = \left[ {\begin{array}{*{20}{c}}
        {{F_{\gamma \gamma }}}&{{F_{\gamma T}}}\\
        {{F_{T\gamma }}}&{{F_{TT}}}
        \end{array}} \right]= \left[ {\begin{array}{*{20}{c}}
    {2{\rm vec}{{\left[ {{\partial _\gamma }\rho } \right]}^T}^{}{\Lambda ^{ - 1}}{\rm vec}\left[ {{\partial _\gamma }\rho } \right]}&{2{\rm vec}{{\left[ {{\partial _\gamma }\rho } \right]}^T}{\Lambda ^{ - 1}}{\rm vec}\left[ {{\partial _T}\rho } \right]}\\
    {2{\rm vec}{{\left[ {{\partial _T}\rho } \right]}^T}{\Lambda ^{ - 1}}{\rm vec}\left[ {{\partial _\gamma }\rho } \right]}&{2{\rm vec}{{\left[ {{\partial _T}\rho } \right]}^T}{\Lambda ^{ - 1}}{\rm vec}\left[ {{\partial _T}\rho } \right]}
    \end{array}} \right].
\end{equation}
It is simple to verify that
\begin{equation}
{F_{\gamma \gamma }} = 4\left[ {\left( {\alpha  + \tau } \right)\left( {{{\left( {{\partial _\gamma }a} \right)}^2} + {{\left( {{\partial _\gamma }x} \right)}^2}} \right) + \left( {\nu  + \omega } \right)\left( {{{\left( {{\partial _\gamma }b} \right)}^2} + {{\left( {{\partial _\gamma }z} \right)}^2}} \right) + 4\xi \,{\partial _\gamma }a\,\,{\partial _\gamma }x + 4\mu \,{\partial _\gamma }b\,\,{\partial _\gamma }z} \right],
\end{equation}
\begin{equation}
{F_{TT}} = 4\left[ {\left( {\alpha  + \tau } \right)\left( {{{\left( {{\partial _T}a} \right)}^2} + {{\left( {{\partial _T}x} \right)}^2}} \right) + \left( {\nu  + \omega } \right)\left( {{{\left( {{\partial _T}b} \right)}^2} + {{\left( {{\partial _T}z} \right)}^2}} \right) + 4\xi \,{\partial _\gamma}a\,\,{\partial _T}x + 4\mu \,{\partial _T}b\,\,{\partial _T}z} \right],
\end{equation}
and
 \begin{align}
{F_{\gamma T}} =&4\left( {\alpha  + \tau } \right)\left( {{\partial _\gamma }a\,\,{\partial _T}a + {\partial _\gamma }x\,\,{\partial _T}x} \right) + 4\left( {\nu  + \omega } \right)\left( {{\partial _\gamma }b\,\,{\partial _T}b + {\partial _\gamma }z\,\,{\partial _T}z} \right) \notag\\& +8\xi \left( {\,{\partial_\gamma}a\,\,{\partial _T}x + \,{\partial _\gamma}x\,\,{\partial _T}a} \right) + 8\mu \,\left( {{\partial _\gamma }b\,\,{\partial _T}z + {\partial _\gamma }z\,\,{\partial _T}b} \right).
\end{align}
The optimal estimator, in any given quantum metrology protocol, is defined as one which saturates the quantum Cramer-Rao inequality. This bound is a lower limit of the covariance matrix of estimators ${\hat \theta } = \left( {\gamma ,T} \right)$ and it reads
\begin{equation}
{\rm{Cov}}\left( \hat \theta \right) \ge {F^{ - 1}}. \label{CR}
\end{equation}
The inverse of quantum Fisher information matrix is given by
\begin{equation}
{F^{ - 1}} = \frac{1}{{\det \left( F \right)}}\left[ {\begin{array}{*{20}{c}}
{{F_{TT}}}&{ - {F_{\gamma T}}}\\
{ - {F_{\gamma T}}}&{{F_{\gamma \gamma }}}
\end{array}} \right].
\end{equation}
Therefore, from the inequality (\ref{CR}), one gets \cite{Prussing1986}
\begin{equation}
{{\rm Var}\left( \gamma\right) \ge \frac{{{F_{TT}}}}{{\det \left( F \right)}}}, \label{varG1}
\end{equation}
\begin{equation}
{\rm Var}\left( T \right) \ge \frac{{{F_{\gamma \gamma }}}}{{\det \left( F \right)}}, \label{labT1}
\end{equation}
and
\begin{equation}
\left( {{\rm Var}\left( \gamma\right) - \frac{{{F_{TT}}}}{{\det \left( F \right)}}} \right)\left( {{\rm Var}\left( T\right)  - \frac{{{F_{\gamma \gamma }}}}{{\det \left( F \right)}}} \right) \ge {\left( {{\rm Cov}\left( {\gamma ,T}\right)  + \frac{{{F_{\gamma T}}}}{{\det \left( F \right)}}} \right)^2}. \label{vartvarg1}
\end{equation}
Using the equation (\ref{vecL}), the matricial forms of the symmetric logarithmic derivatives, in term of the parameters $\gamma$ and $T$, are given by
\begin{equation}
{L_\gamma } = 2\left[ {\begin{array}{*{20}{c}}
    {\left( {\alpha  + \tau } \right){\partial _\gamma }\,a + 2\xi {\partial _\gamma }\,x}&0&0&{\left( {\alpha  + \tau } \right){\partial _\gamma }\,x + 2\xi \,{\partial _\gamma }\,a}\\
    0&{(\nu  + \omega ){\mkern 1mu} {\partial _\gamma }\,b + 2\mu \,{\mkern 1mu} {\partial _\gamma }z}&{(\nu  + \omega ){\mkern 1mu} {\partial _\gamma }\,z + 2\mu \,{\mkern 1mu} {\partial _\gamma }b}&0\\
    0&{(\nu  + \omega ){\mkern 1mu} {\partial _\gamma }\,z + 2\mu \,{\mkern 1mu} {\partial _\gamma }b}&{(\nu  + \omega ){\mkern 1mu} {\partial _\gamma }\,b + 2\mu \,{\mkern 1mu} {\partial _\gamma }z}&0\\
    {\left( {\alpha  + \tau } \right){\partial _\gamma }\,x + 2\xi \,{\partial _\gamma }\,a}&0&0&{\left( {\alpha  + \tau } \right){\partial _\gamma }\,a + 2\xi {\partial _\gamma }\,x}
    \end{array}} \right],
\end{equation}
and
\begin{equation}
{L_T} = 2\left[ {\begin{array}{*{20}{c}}
    {\left( {\alpha  + \tau } \right){\partial _T}\,a + 2\xi {\partial _T}\,x}&0&0&{\left( {\alpha  + \tau } \right){\partial _T}\,x + 2\xi \,{\partial _T}\,a}\\
    0&{(\nu  + \omega ){\mkern 1mu} {\partial _T}\,b + 2\mu \,{\mkern 1mu} {\partial _T}z}&{(\nu  + \omega ){\mkern 1mu} {\partial _T}\,z + 2\mu \,{\mkern 1mu} {\partial _T}b}&0\\
    0&{(\nu  + \omega ){\mkern 1mu} {\partial _T}\,z + 2\mu \,{\mkern 1mu} {\partial _T}b}&{(\nu  + \omega ){\mkern 1mu} {\partial _T}\,b + 2\mu \,{\mkern 1mu} {\partial _T}z}&0\\
    {\left( {\alpha  + \tau } \right){\partial _T}\,x + 2\xi \,{\partial _T}\,a}&0&0&{\left( {\alpha  + \tau } \right){\partial _T}\,a + 2\xi {\partial _T}\,x}
    \end{array}} \right].
\end{equation}
The eigenvectors of ${L_\gamma }$ and ${L_T }$ can be expressed as a linear combination of Bell states $\left| {{\psi ^ \pm }} \right\rangle  = \frac{1}{{\sqrt 2 }}\left( {\left| {01} \right\rangle  \pm \left| {10} \right\rangle } \right)$ and $\left| {{\chi ^ \pm }} \right\rangle  = \frac{1}{{\sqrt 2 }}\left( {\left| {00} \right\rangle  \pm \left| {11} \right\rangle } \right)$ which are the eigenstates of the Hamiltonian under consideration (\ref{HN}). They provide the optimal measurement bases such that the limits imposed by the inequalities (\ref{varG1}), (\ref{labT1}) and (\ref{vartvarg1}) are fulfilled. The optimal bases for $\gamma$ and $T$ are given by:
\begin{equation}
{{\bf{B}}_\gamma } = {{\bf{B}}_T} = \left\{ { - \left| {{\psi ^ - }} \right\rangle ,\left| {{\psi ^ + }} \right\rangle , - \left| {{\chi ^ - }} \right\rangle ,\left| {{\chi ^ + }} \right\rangle } \right\}.
\end{equation}
The fact that we have the same optimal estimation bases means that the symmetric logarithmic derivatives ${L_\gamma}$ and ${L_T}$ commute. This will allow us to satisfy and saturate the bounds given by (\ref{varG1}), (\ref{labT1}) and (\ref{vartvarg1}). The saturation of the first two inequalities gives the highest precision on the estimation of the parameters $\gamma$ and $T$. The minimal values of ${\rm Var}{\left( \gamma  \right)}$ and ${\rm Var}{\left(T  \right)}$ are given by
\begin{equation}
{\rm Var}{\left( \gamma  \right)_{\min }} = {{T^2}\left( 1 + {\gamma ^2} + \left( {1 + {\gamma ^2}} \right)\cosh\left( \beta \right) \cosh\left( \beta\gamma \right) - 2\gamma \sinh\left( \beta\right)\sinh\left( \beta\gamma\right) \right)},
\end{equation}

\begin{equation}
{\rm{Var}}{\left(T \right)_{\min }} = {T^4}\left[ {\frac{3}{2} + \frac{{\cosh \left( {\beta \left( {\gamma  - 2} \right)} \right) + \cosh \left( {\beta \left( {2 + \gamma } \right)} \right) + \cosh \left( {\beta \left( {1 - 2\gamma } \right)} \right) + \cosh \left( {\beta \left( {1 + 2\gamma } \right)} \right)}}{{4\left( {\cosh \left( \beta  \right) + \cosh \left( {\beta \gamma } \right)} \right)}}} \right].
\end{equation}

\begin{figure}[H]
    \centering
    \begin{minipage}[t]{3in}
        \centering
        \includegraphics[width=3in]{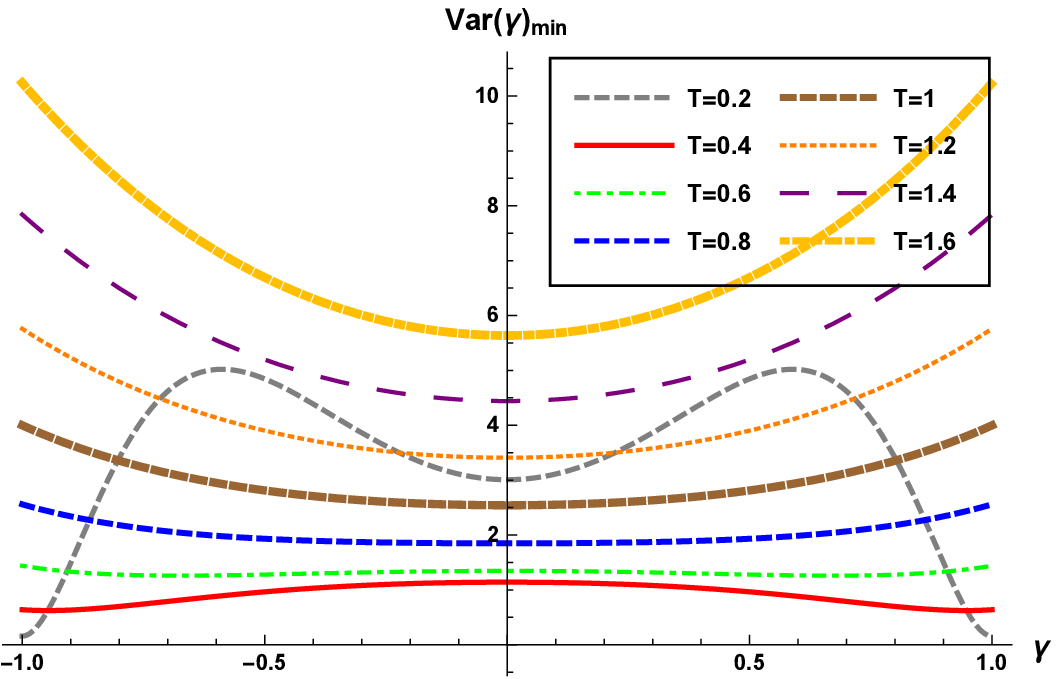}
    \end{minipage} \hspace{0.5cm}
    \begin{minipage}[t]{3in}
        \centering
        \includegraphics[width=3in]{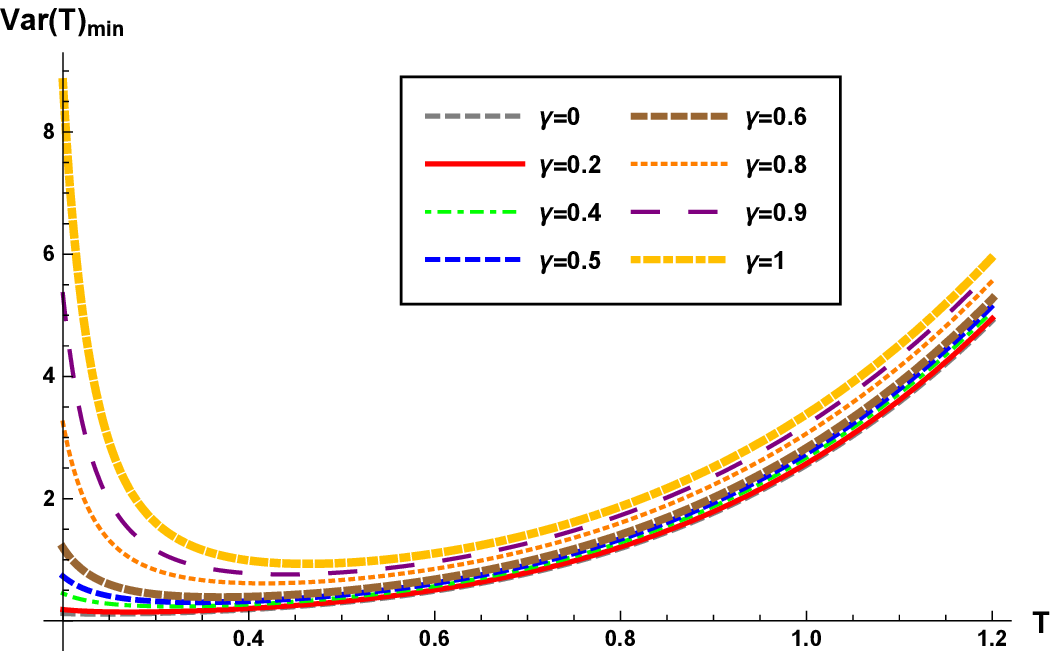}
    \end{minipage}

    \caption{\sf The minimal variances of simultaneous estimates of parameters $\gamma$ and $T$ }
    \label{Fig1}
\end{figure}
The results reported in Fig.(\ref{Fig1}) (the left panel) represent the minimal variance of the estimator of the parameter $\gamma$ estimated simultaneously. These results show that at low temperature, the highest precision of the parameter $\gamma$ is ${\gamma _{opt}} = \pm1$ which corresponds to Ising model. On the other hand, best precision of $\gamma$ with high temperature corresponds to the case where ${\gamma _{opt}} = 0$. This corresponds to the isotropic $XX$ model. The right panel of Fig.(\ref{Fig1}) represents the evolution of the minimal variance of the estimator of the temperature estimated simultaneously. This variance shows that the optimal value of the temperature $T$ is almost equal to ${T_{opt}} = 0.25$ for the case of the isotropic $XX$ model and ${T_{opt}} = 0.5$ for Ising model.\\
Now, we consider the situation in which we estimate the parameters individually. For this, we assume that the parameters are statistically independent. This means that the precise identification of a single parameter does not affect the accuracy of others. This is only true in the case where $F_{ij}=0$ $(i \neq j)$. This implies
\begin{equation}
{\rm Var}\left( \gamma  \right)^{\rm Ind} \ge F_{\gamma \gamma }^{ - 1}, \hspace{1.5cm} {\rm Var}\left( T  \right)^{\rm Ind} \ge F_{TT }^{ - 1} \label{varind}.
\end{equation}
The saturation of these last two inequalities leads to
{\small\begin{equation}
{\rm Var}\left( \gamma \right)_{\min }^{\rm Ind} = \frac{{4{T^2}{{\left( \cosh\left(\beta \right) + \cosh \left( {\beta \gamma } \right) \right)}^3}}}{{6\left( {\cosh \left( \beta  \right) + \cosh \left( {\beta \gamma } \right)} \right) + \cosh \left( \beta \left( {\gamma  - 2} \right) \right) + \cosh\left(\beta \left( {\gamma  + 2} \right) \right) + \cosh \left( \beta \left( {1 - 2\gamma } \right) \right) + \cosh \left( \beta \left( 1 + 2\gamma  \right) \right)}},
\end{equation}}
and
\begin{equation}
{\rm Var}\left( T  \right)_{\min }^{\rm Ind} =\frac{{{T^4}{{\left( {\cosh \left( \beta  \right) + \cosh \left(\beta \gamma \right)} \right)}^2}}}{{\left( 1 + \gamma ^2 \right)\left( 1 + \cosh \left( \beta  \right)\cosh \left(\beta \gamma\right) \right) - 2\gamma \sinh \left( \beta  \right)\sinh\left( \beta \gamma \right)}}.
\end{equation}
\begin{figure}[H]
    \centering
    \begin{minipage}[t]{3in}
        \centering
        \includegraphics[width=3in]{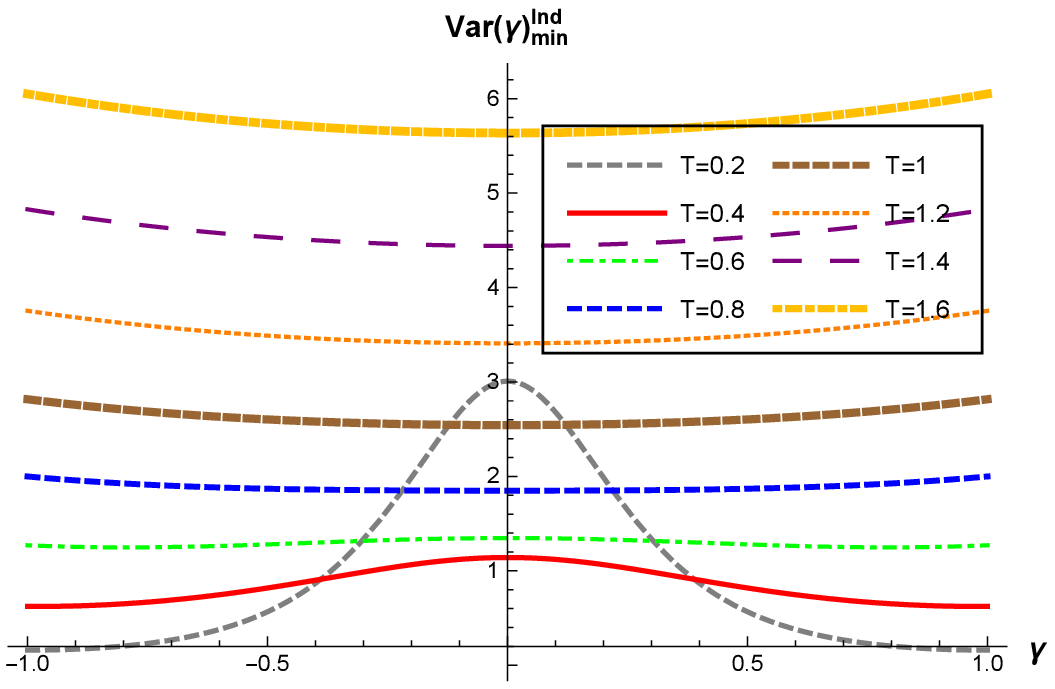}
    \end{minipage} \hspace{0.5cm}
    \begin{minipage}[t]{3in}
        \centering
        \includegraphics[width=3in]{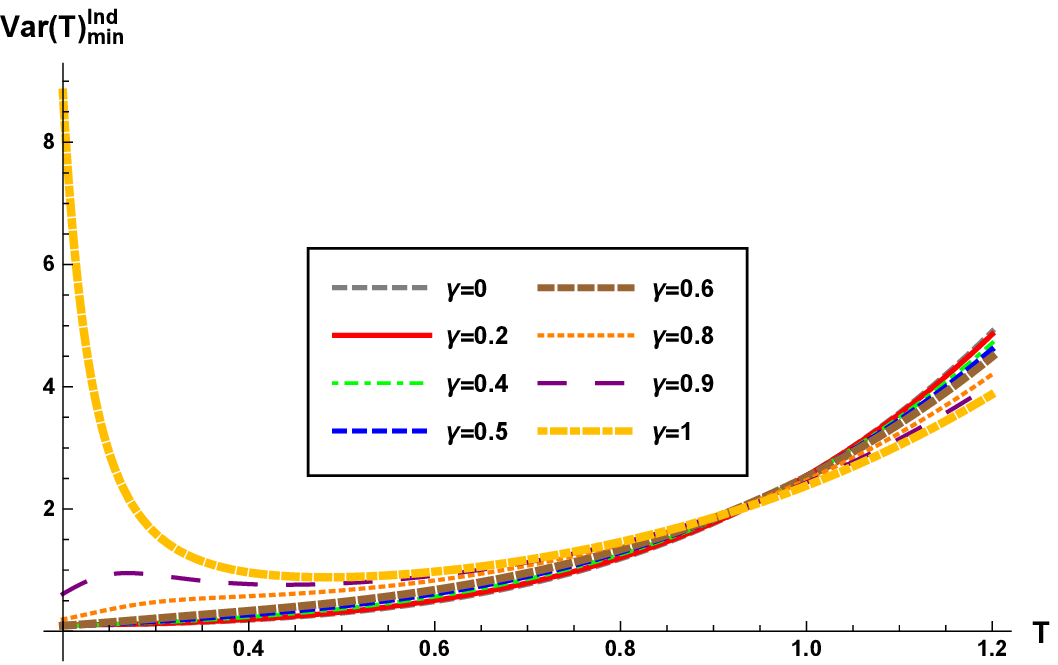}
    \end{minipage}

    \caption{\sf The minimal variances of individual estimates of parameters $\gamma$ and $T$ }
    \label{Fig2}
\end{figure}
The evolution of the minimal variances of individual estimation of the parameters $\gamma$ and $T$ is depicted in Fig.(\ref{Fig2}). We note that the obtained behavior is almost similar to that observed in the strategy of simultaneous estimation. But there is some difference in the uncertainty concerning the precise estimation depending on the type of strategy we employed in the measurement. To compare the obtained results in both cases, we introduce the ratio between the total variance in the individual and simultaneous schemes. It is defined as follows:
\begin{equation}
\Gamma  = \frac{{{\Delta _{\rm Sim}}}}{{{\Delta_{\rm Ind}}}}, \label{Raport}
\end{equation}
with ${\Delta_{\rm Ind}} = {\rm{Var}}_{\min }^{\rm Ind}\left( \gamma  \right) + {\rm{Var}}_{\min }^{\rm Ind}\left( T \right)$ and ${\Delta_{\rm Sim}} = \frac{1}{2}\left( {{\rm{Var}}{{\left( \gamma  \right)}_{\min }} + {\rm{Var}}{{\left( T \right)}_{\min }}} \right)$. After some simplifications, we obtain
\begin{equation}
\Gamma  = \frac{{\left( {1 + \cosh \left( \beta  \right)\cosh \left( {\beta \gamma } \right)} \right)\left( {\left( {1 + {\gamma ^2}} \right)\left( {1 + \cosh \left( \beta  \right)\cosh \left( {\beta \gamma } \right)} \right) - 2\gamma \sinh \left( \beta  \right)\sinh \left( {\beta \gamma } \right)} \right)}}{{2{{\left( {\cosh \left( \beta  \right) + \cosh \left( {\beta \gamma } \right)} \right)}^2}}}. \label{Rapport1}
\end{equation}
\begin{figure}[H]
    \begin{centering}
\includegraphics{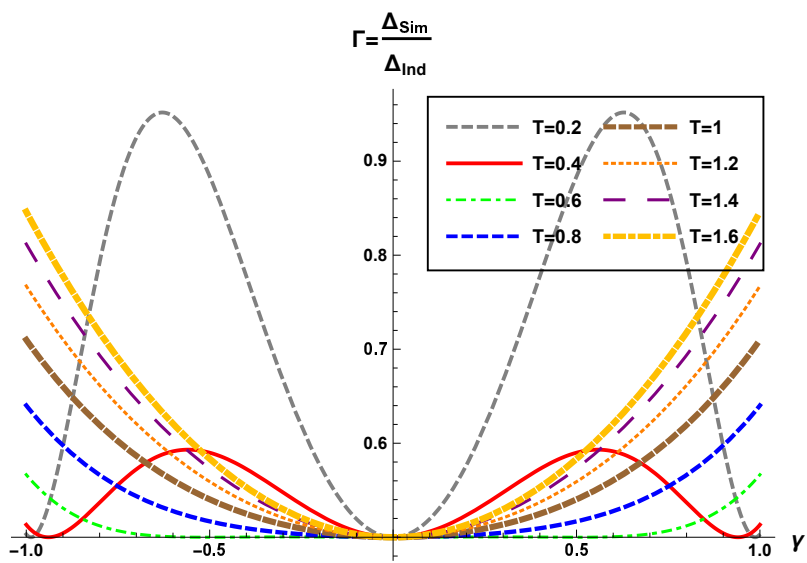}
\caption{\sf The ratio between the minimal total variances in estimating the parameters $\gamma$ and $T$.}
\label{Fig3}
    \end{centering}
\end{figure}
In order to assess the performance of a metrological strategy, the ratio $\Gamma$ (eq.(\ref{Rapport1})) is plotted in Fig.(\ref{Fig3}). As it can be seen from Fig.(\ref{Fig3}), we have $\Gamma\leq1$ (i.e., $\Delta _{\rm Sim} \leq \Delta _{\rm Ind}$). This clearly shows that the simultaneous estimation strategy offers an advantage in the context of improving the precision in comparison with the individual estimation scheme.
\subsection{Isotropic XY model with a magnetic field}
Now we consider the two qubit isotropic $XY$ model ($N=2$, $J_x=J_y=J$ and $J_z=0$) subjected to a external magnetic field $B$ which is described by the following Hamiltonian
\begin{equation}
    H = \sum\limits_{n = 1}^2 {J\left( {S_n^xS_{n + 1}^x + S_n^yS_{n + 1}^y} \right)}  + B\sum\limits_{n = 1}^2 {S_n^z}.
\end{equation}
It rewrites, in terms of the raising and lowering operators ${\sigma _n^ \pm}$, as
\begin{equation}
    H = \frac{B}{2}\left( {\sigma _1^z + \sigma _2^z} \right) + J\left( {\sigma _1^ + \sigma _2^ -  + \sigma _2^ + \sigma _1^ - } \right).
\end{equation}
The Hamiltonian $H$ satisfies the following eigenvalues equations
\begin{eqnarray}
H\left| {00} \right\rangle  = B\left| {00} \right\rangle ;\hspace{0.5cm}H\left| {11} \right\rangle  =  - B\left| {11} \right\rangle ;\hspace{0.5cm}H\left| {{\psi ^ \pm }} \right\rangle  = B\left| {{\psi ^ \pm }} \right\rangle,
\end{eqnarray}
where $\left| {{\psi ^ \pm }} \right\rangle $ are the Bell states defined by $\left| {{\psi ^ \pm }} \right\rangle  = {{\left( {\left| {00} \right\rangle  \pm \left| {11} \right\rangle } \right)} \mathord{\left/
        {\vphantom {{\left( {\left| {00} \right\rangle  \pm \left| {11} \right\rangle } \right)} {\sqrt 2 }}} \right.
        \kern-\nulldelimiterspace} {\sqrt 2 }}$. The density matrix $\rho  = {{{\exp\left( - \beta H\right)}} \mathord{\left/
        {\vphantom {{{e^{ - \beta H}}} Z}} \right.
        \kern-\nulldelimiterspace} Z}$ of this system is given by
\begin{equation}
\rho  = \left[ {\begin{array}{*{20}{c}}
    c&0&0&0\\
    0&t&y&0\\
    0&y&t&0\\
    0&0&0&d
    \end{array}} \right]. \label{gho2case}
\end{equation}
where
\begin{equation}
    c = \frac{{{e^{ - \beta B}}}}{{2\left( {\cosh\left( \beta B \right) + \cosh\left( \beta J \right)} \right)}}, \hspace{1.5cm}d = \frac{{{e^{\beta B}}}}{{2\left( {\cosh\left( \beta B \right) + \cosh\left( \beta J \right)} \right)}},
\end{equation}
 \begin{equation}
    y = \frac{{ - \sinh\left( \beta J \right)}}{{2\left( {\cosh\left( \beta B \right) + \cosh\left( \beta J \right)} \right)}},\hspace{1.5cm}t = \frac{{\cosh\left( \beta J \right)}}{{2\left( {\cosh\left( \beta B \right) + \cosh\left( \beta J \right)} \right)}}.
 \end{equation}
For this two-qubit system, we consider the estimation of the magnetic field $B$ and the temperature $T$ (i.e., ${\hat \theta } \equiv \left( {B ,T} \right)$). In this case, the matrix $\Lambda$ (of Eq.(\ref{landa})) is given by
\begin{equation}
\Lambda  = \left[ {\begin{array}{*{20}{c}}
    {{\Lambda _{11}}}&{{0_{4 \times 4}}}&{{0_{4 \times 4}}}&{{0_{4 \times 4}}}\\
    {{0_{4 \times 4}}}&{{\Lambda _{22}}}&{{\Lambda _{23}}}&{{0_{4 \times 4}}}\\
    {{0_{4 \times 4}}}&{{\Lambda _{32}}}&{{\Lambda _{33}}}&{{0_{4 \times 4}}}\\
    {{0_{4 \times 4}}}&{{0_{4 \times 4}}}&{{0_{4 \times 4}}}&{{\Lambda _{44}}}
    \end{array}} \right],\label{LLL}
\end{equation}
with
\begin{equation}
{\Lambda _{11}} = \left[ {\begin{array}{*{20}{c}}
    {2c}&0&0&0\\
    0&{c + t}&y&0\\
    0&y&{c + t}&0\\
    0&0&0&{d + c}
    \end{array}} \right], \hspace{1cm}{\Lambda _{22}} = {\Lambda _{33}} = \left[ {\begin{array}{*{20}{c}}
{c + t}&0&0&0\\
0&{2t}&y&0\\
0&y&{2t}&0\\
0&0&0&{c + t}
\end{array}} \right],
\end{equation}
and
\begin{equation}
{\Lambda _{23}} = {\Lambda _{32}} = \left[ {\begin{array}{*{20}{c}}
    y&0&0&0\\
    0&y&0&0\\
    0&0&y&0\\
    0&0&0&y
    \end{array}} \right], \hspace{1cm}{\Lambda _{44}} = \left[ {\begin{array}{*{20}{c}}
{d + c}&0&0&0\\
0&{c + t}&y&0\\
0&y&{c + t}&0\\
0&0&0&{2d}
\end{array}} \right].
\end{equation}
The inverse of matrix $\Lambda$ (\ref{LLL}) takes the form
\begin{equation}
{\Lambda ^{ - 1}} = \left[ {\begin{array}{*{20}{c}}
    {{{\left( {{\Lambda ^{ - 1}}} \right)}_{11}}}&{{0_{4 \times 4}}}&{{0_{4 \times 4}}}&{{0_{4 \times 4}}}\\
    {{0_{4 \times 4}}}&{{{\left( {{\Lambda ^{ - 1}}} \right)}_{22}}}&{{{\left( {{\Lambda ^{ - 1}}} \right)}_{23}}}&{{0_{4 \times 4}}}\\
    {{0_{4 \times 4}}}&{{{\left( {{\Lambda ^{ - 1}}} \right)}_{32}}}&{{{\left( {{\Lambda ^{ - 1}}} \right)}_{33}}}&{{0_{4 \times 4}}}\\
    {{0_{4 \times 4}}}&{{0_{4 \times 4}}}&{{0_{4 \times 4}}}&{{{\left( {{\Lambda ^{ - 1}}} \right)}_{44}}}
    \end{array}} \right],
\end{equation}
where
\begin{equation}
{\left( {{\Lambda ^{ - 1}}} \right)_{11}} = \left[ {\begin{array}{*{20}{c}}
    n&0&0&0\\
    0&p&r&0\\
    0&r&p&0\\
    0&0&0&s
    \end{array}} \right], \hspace{1cm}{\left( {{\Lambda ^{ - 1}}} \right)_{22}} = {\left( {{\Lambda ^{ - 1}}} \right)_{33}} = \left[ {\begin{array}{*{20}{c}}
p&0&0&0\\
0&e&f&0\\
0&f&e&0\\
0&0&0&g
\end{array}} \right],
\end{equation}
and
\begin{equation}
{\left( {{\Lambda ^{ - 1}}} \right)_{44}} = \left[ {\begin{array}{*{20}{c}}
    s&0&0&0\\
    0&g&l&0\\
    0&l&g&0\\
    0&0&0&m
    \end{array}} \right],\hspace{1cm}{\left( {{\Lambda ^{ - 1}}} \right)_{23}} = {\left( {{\Lambda ^{ - 1}}} \right)_{32}} = \left[ {\begin{array}{*{20}{c}}
h&0&0&0\\
0&f&k&0\\
0&k&f&0\\
0&0&0&l
\end{array}} \right],
\end{equation}
with
\begin{equation}
    n = {e^{\beta B}}\left( {\cosh\left( \beta B \right) + \cosh\left( \beta J \right)} \right),\hspace{1.5cm}p = 1 + {e^{\beta B}}\cosh\left( \beta J \right),
\end{equation}
\begin{equation}
      r = {e^{\beta B}}\sinh\left( \beta J \right), \hspace{1cm} s = 1 + \frac{{\cosh\left( \beta J \right)}}{{\cosh\left( \beta B \right)}},\hspace{1cm} l = \left( {\cosh\left( \beta B \right) - \sinh\left( \beta B \right)} \right)\sinh\left( \beta J \right),
\end{equation}
\begin{equation}
    e = \frac{1}{4}\left( {\cosh\left( \beta B \right) + \cosh\left( \beta J \right)} \right)\left( {3 + \cosh\left( 2\beta J \right)} \right){\mathop{\rm sech}\nolimits} \left( \beta J \right),
\end{equation}
\begin{equation}
    f = \frac{1}{2}\left( {\cosh\left( \beta B \right) + \cosh\left( \beta J \right)} \right){\mathop{\rm \sinh}\nolimits} \left( \beta J \right), \hspace{1.5cm}g = 1 + {e^{ - \beta B}}\cosh\left( \beta J \right),
\end{equation}
\begin{equation}
    h = {e^{\beta B}}\sinh\left( \beta J \right), \hspace{1.5cm}k = \frac{1}{2}\left( {\cosh\left( \beta B \right) + \cosh\left( \beta J \right)} \right)\tanh\left( \beta J \right)\sinh\left( \beta J \right),
\end{equation}
The vec-operator associated the density matrix derivatives, with respect to parameters $B$ and $T$, are given by
\begin{eqnarray}
{\rm vec}\left[ {{\partial _B}\rho } \right] = {\left( {{\partial _B}c,0,0,0,0,{\partial _B}t,{\partial _B}y,0,0,{\partial _B}y,{\partial _B}t,0,0,0,0,{\partial _B}d} \right)^T},
\end{eqnarray}
and
\begin{eqnarray}
{\rm vec}\left[ {{\partial _T}\rho } \right] = {\left( {{\partial _T}c,0,0,0,0,{\partial _T}t,{\partial _T}y,0,0,{\partial _T}y,{\partial _T}t,0,0,0,0,{\partial _T}d} \right)^T}.
\end{eqnarray}
The quantum Fisher information matrix writes
\begin{eqnarray}
F = \left[ {\begin{array}{*{20}{c}}
    {2{\rm vec}{{\left[ {{\partial _B}\rho } \right]}^T}^{}{\Lambda ^{ - 1}}{\rm vec}\left[ {{\partial _B}\rho } \right]}&{2{\rm vec}{{\left[ {{\partial _B}\rho } \right]}^T}{\Lambda ^{ - 1}}{\rm vec}\left[ {{\partial _T}\rho } \right]}\\
    {2{\rm vec}{{\left[ {{\partial _T}\rho } \right]}^T}{\Lambda ^{ - 1}}{\rm vec}\left[ {{\partial _B}\rho } \right]}&{2{\rm vec}{{\left[ {{\partial _T}\rho } \right]}^T}{\Lambda ^{ - 1}}{\rm vec}\left[ {{\partial _T}\rho } \right]}
    \end{array}} \right].
\end{eqnarray}
After a straightforward calculation, the elements of the quantum Fisher information matrix are obtained analytically as
\begin{equation}
{F_{BB}} = \frac{{2{{\rm{e}}^{\beta B}}\left( {2{{\rm{e}}^{\beta B}} + \left( {1 + {{\rm{e}}^{2\beta B}}} \right)\cosh\left( \beta J\right) } \right)}}{{{T^2}{{\left( {1 + {{\rm{e}}^{2\beta B}} + 2{{\rm{e}}^{\beta B}}\cosh\left( \beta J\right) } \right)}^2}}},
\end{equation}
\begin{equation}
{F_{BT}} = {F_{TB}} = \frac{{2{{\rm{e}}^{\beta B}}\left( {2B{{\rm{e}}^{\beta B}} + B\left( {1 + {{\rm{e}}^{2\beta B}}} \right)\cosh\left(  \beta J \right)  - \left( { - 1 + {{\rm{e}}^{2\beta B}}} \right)J\sinh \left(  \beta J \right) } \right)}}{{{T^3}{{\left( {1 + {{\rm{e}}^{2\beta B}} + 2{{\rm{e}}^{\beta B}}\cosh\left( \beta J \right) } \right)}^2}}},
\end{equation}

\begin{equation}
{F_{TT}} = \frac{{{{\rm{e}}^{ - 2\beta B}}\left( {1 + {{\rm{e}}^{2\beta B}} + 2{{\rm{e}}^{\beta B}}\cosh \left( \beta J \right)} \right)}}{{4{T^4}{{\left( {\cosh\left( \beta B \right) + \cosh\left( \beta J \right)} \right)}^3}}}\left( \begin{array}{l}
\left( {1 + {{\rm{e}}^{2\beta B}}} \right)\left( {{B^2} + {J^2}} \right)\cosh\left( \beta J\right) + \\
2\left( {{{\rm{e}}^{\beta B}}\left( {{B^2} + {J^2}} \right) - B\left( { - 1 + {{\rm{e}}^{2\beta B}}} \right)J\sinh\left( \beta J \right)} \right)
\end{array} \right).
\end{equation}
The inverse of the quantum Fisher information matrix is given by
\begin{equation}
{F^{ - 1}} = \frac{1}{{\det \left( F \right)}}\left[ {\begin{array}{*{20}{c}}
    {{F_{TT}}}&{ - {F_{BT}}}\\
    { - {F_{BT}}}&{{F_{BB}}}
    \end{array}} \right].
\end{equation}
The equation (\ref{CR}) gives, in this case, the following inequalities
\begin{eqnarray}
{\rm Var}\left( B\right) \ge \frac{{{F_{TT}}}}{{\det \left( F \right)}}, \label{varb2}
\end{eqnarray}
\begin{eqnarray}
{\rm Var}\left( T\right) \ge \frac{{{F_{BB}}}}{{\det \left( F \right)}}, \label{vart2}
\end{eqnarray}
and
\begin{eqnarray}
\left( {{\rm Var}\left( B\right) - \frac{{{F_{TT}}}}{{\det \left( F \right)}}} \right)\left( {{\rm Var}\left( T\right)  - \frac{{{F_{BB}}}}{{\det \left( F \right)}}} \right) \ge {\left( {{\rm Cov}\left( B,T\right)  + {F_{BT}}} \right)^2}. \label{vartvarb2}
\end{eqnarray}
Using the equation (\ref{vecL}), the operators of the symmetric logarithmic derivative ${L_B}$ et ${L_T}$ are respectively given by
\begin{equation}
{L_B} = 2\left[ {\begin{array}{*{20}{c}}
    {n\,{\partial _B}\,c}&0&0&0\\
    0&{(e + k){\mkern 1mu} {\partial _B}\,t + 2f{\mkern 1mu} {\partial _B}y}&{(e + k){\mkern 1mu} {\partial _B}y + 2f{\mkern 1mu} {\partial _B}t}&0\\
    0&{(e + k){\mkern 1mu} {\partial _B}y + 2f{\mkern 1mu} {\partial _B}t}&{(e + k){\mkern 1mu} {\partial _B}t + 2f{\mkern 1mu} {\partial _B}y}&0\\
    0&0&0&{m{\mkern 1mu} {\partial _B}d}
    \end{array}} \right],
\end{equation}
\begin{equation}
{L_T} = 2\left[ {\begin{array}{*{20}{c}}
    {n\,{\partial _T}\,c}&0&0&0\\
    0&{(e + k){\mkern 1mu} {\partial _T}\,t + 2f{\mkern 1mu} {\partial _T}y}&{(e + k){\mkern 1mu} {\partial _T}y + 2f{\mkern 1mu} {\partial _T}t}&0\\
    0&{(e + k){\mkern 1mu} {\partial _T}y + 2f{\mkern 1mu} {\partial _T}t}&{(e + k){\mkern 1mu} {\partial _T}t + 2f{\mkern 1mu} {\partial _T}y}&0\\
    0&0&0&{m{\mkern 1mu} {\partial _T}d}
    \end{array}} \right].
\end{equation}
The eigenvectors of the operators ${L_B}$ and ${L_T}$ give the optimal measurement bases that will allow us to reach the bounds in the inequalities (\ref{varb2}), (\ref{vart2}) and (\ref{vartvarb2}). It is simple to verify that the optimal measurement basis is
\begin{equation}
{{\bf{B}}_B} = {{\bf{B}}_T} = \left\{ {\left| {00} \right\rangle ,\left| {{\psi ^ + }} \right\rangle , - \left| {{\psi ^ - }} \right\rangle ,\left| {11} \right\rangle } \right\}.
\end{equation}
The symmetric logarithmic derivatives ${L_B}$ and ${L_T }$ commute and a common eigenbasis can be constructed using the eigenvectors of the Hamiltonian. This basis is the optimal estimation basis to estimate the magnetic field $B$ and the temperature $T$. The analytical expressions of the minimum variances that give the highest precision for the estimation of parameters $B$ and $T$ are
\begin{equation}
    {\rm{Var}}{\left(B\right)_{\min }} = \frac{{{{\rm{e}}^{ - 4\beta B}}{T^2}{{\left( 1 + {{\rm{e}}^{2\beta B}} + 2{{\rm{e}}^{\beta B}}\cosh \left( \beta J \right) \right)}^3}}}{{16{J^2}{{\left( {\cosh \left( \beta B \right) + \cosh\left(\beta J\right)} \right)}^3}}}\left( \begin{array}{l}
    \left( {1 + {{\rm{e}}^{2\beta B}}} \right)\left( {{B^2} + {J^2}} \right)\cosh\left( \beta J\right) + \\
    2\left( {{{\rm{e}}^{\beta B}}\left( {{B^2} + {J^2}} \right) - B\left( { - 1 + {{\rm{e}}^{2\beta B}}} \right)J\sinh\left(\beta J \right)} \right)
    \end{array} \right), \label{varminB}
\end{equation}
\begin{equation}
{\mathop{\rm Var}}\left(T\right)_{\min } = \frac{{{{\rm{e}}^{ - \beta B}}{T^4}\left( {2{{\rm{e}}^{\beta B}} + \left( {1 + {{\rm{e}}^{2\beta B}}} \right)\cosh\left( \beta J\right)} \right)}}{{2{J^2}}}. \label{varminT2}
\end{equation}
According to the equations above (eqs.(\ref{varminB}) and (\ref{varminT2})), it is easy to show that whatever the system is, antiferromagnetic or ferromagnetic (i.e., whatever the value of $J$ is positive or negative),  the behaviors of the minimal variances of the estimators of $B$ and $T$ remain unchanged.
\begin{figure}[H]
    \centering
    \begin{minipage}[t]{3in}
        \centering
        \includegraphics[width=3.1in]{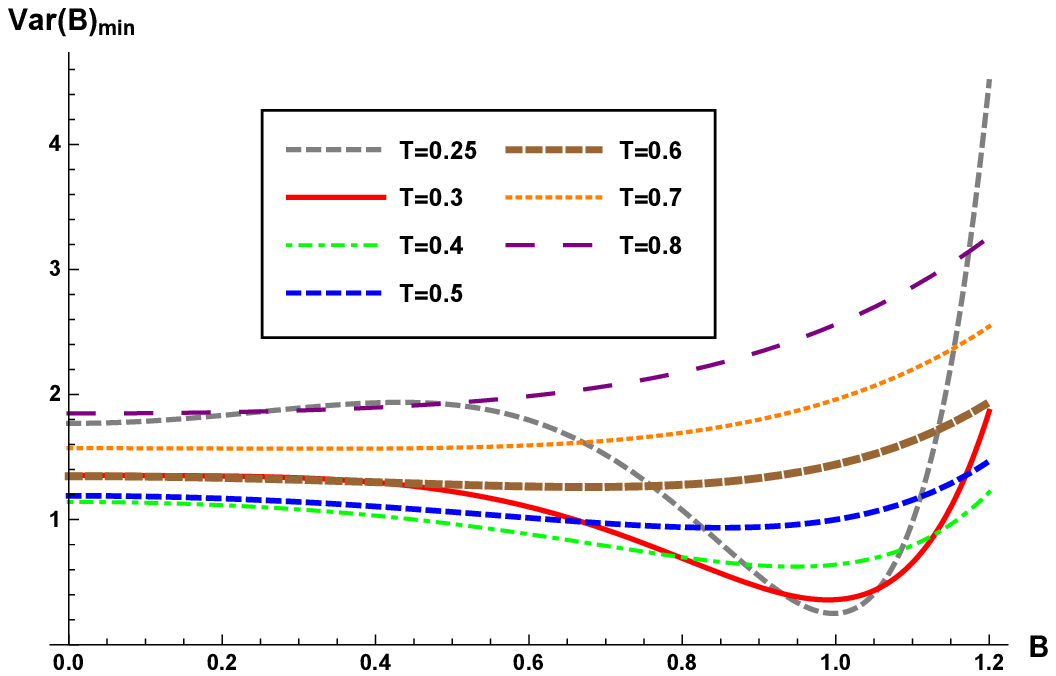}
    \end{minipage}
\hspace{0.5cm}
    \begin{minipage}[t]{3in}
        \centering
        \includegraphics[width=3.1in]{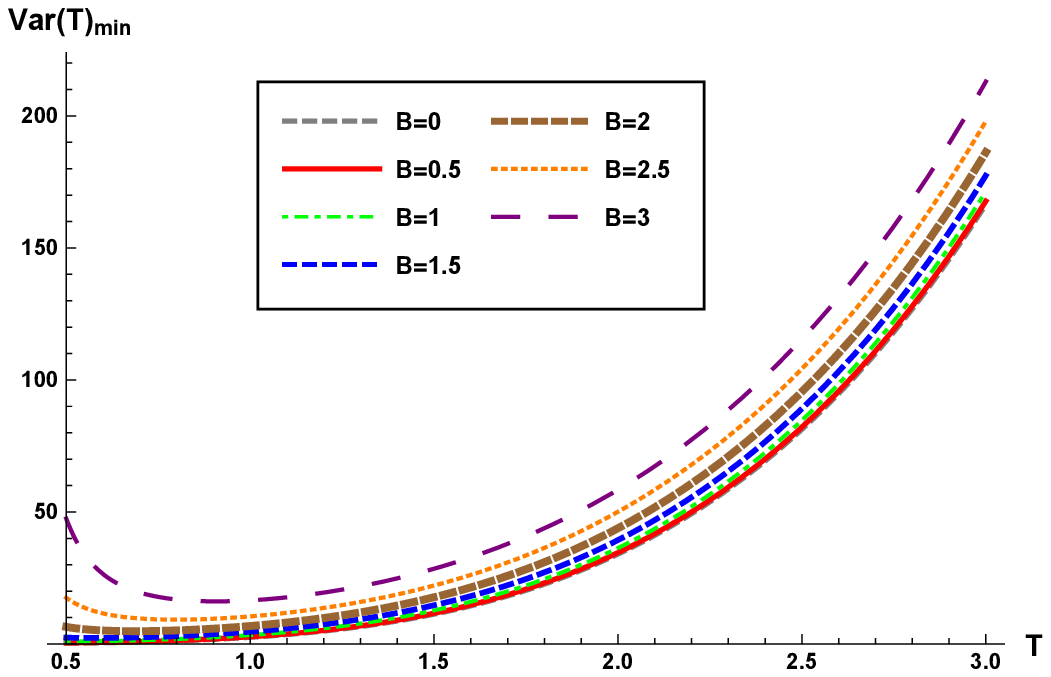}
    \end{minipage}

     \caption{\sf The variances of simultaneous estimates of parameters $B$ and $T$ with $J=1$.}
    \label{fig4}
\end{figure}
Fig.(\ref{fig4}) shows the results of the minimal variances of simultaneous estimation of the magnetic field $B$ and temperature $T$ for $J=1$. For low temperatures, the optimal value of the parameter $B$ is $B_{opt}=1$. The variance of $B$ is minimal, for high temperature, when $B_{opt}=0$. This implies that the isotropic $XY$ model at high temperature has optimal states when the magnetic field is absent. On the other hand, we remark that the variance of the temperature is minimal for $I_c = \left[ 0.5,0.7 \right]$. This interval is usually called the confidence interval in quantum metrology. Now, if we estimate the parameters $B$ and $T$ individually, the Cramer-Rao inequality writes
\begin{equation}
{\rm Var}{\left( B \right)^{Ind}} \ge F_{BB}^{ - 1} \hspace{1.5cm} {\rm Var}{\left( T \right)^{Ind}} \ge F_{TT}^{ - 1},
\end{equation}
with
\begin{equation}
{\rm Var}\left( B \right)_{\min }^{Ind} = \frac{{{T^2}{e^{ - \beta B}}{{\left( {1 + {e^{2\beta B}} + 2{e^{\beta B}}\cosh \left( {\beta J} \right)} \right)}^2}}}{{2\left( {2{e^{\beta B}} + \left( {1 + {e^{2\beta B}}} \right)\cosh \left( {\beta J} \right)} \right)}}, \label{varminindB}
\end{equation}
and
\begin{equation}
{\rm Var}\left( T \right)_{\min }^{Ind} = \frac{{{T^4}{{\left( {\cosh \left( {\beta B} \right) + \cosh \left( {\beta J} \right)} \right)}^2}}}{{\left( {{B^2} + {J^2}} \right)\left( {1 + \cosh \left( {\beta B} \right)\cosh \left( {\beta J} \right)} \right) - 2BJ\sinh \left( {\beta B} \right)\sinh \left( {\beta J} \right)}}. \label{varminindT2}
\end{equation}

\begin{figure}[H]
    \centering
    \begin{minipage}[t]{3in}
        \centering
        \includegraphics[width=3.1in]{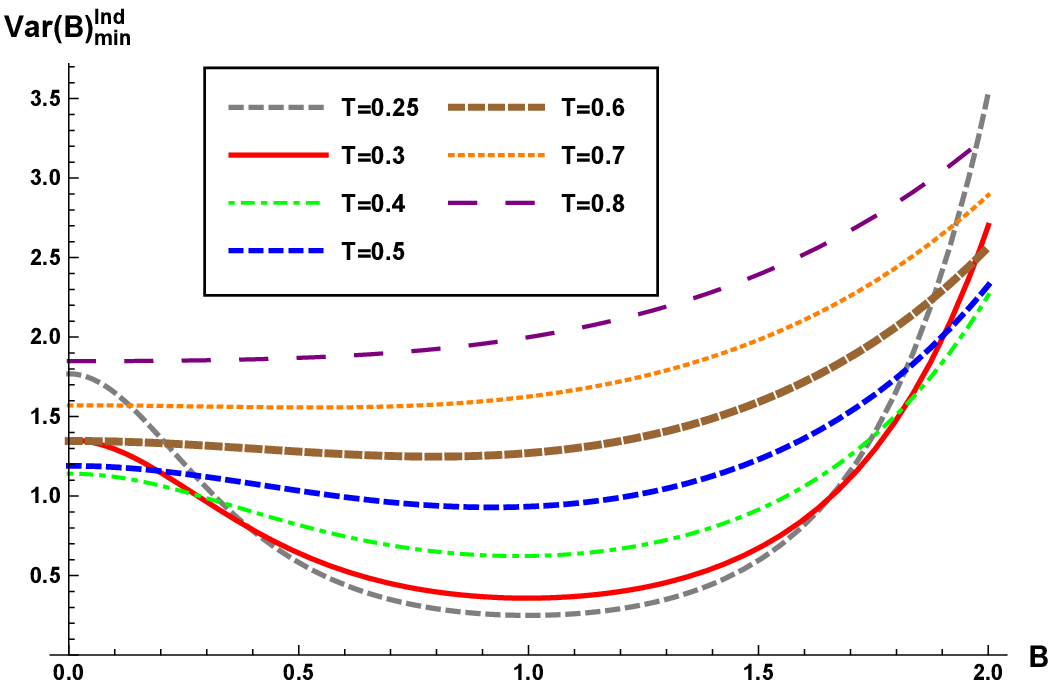}
    \end{minipage}
    \hspace{0.5cm}
    \begin{minipage}[t]{3in}
        \centering
        \includegraphics[width=3.1in]{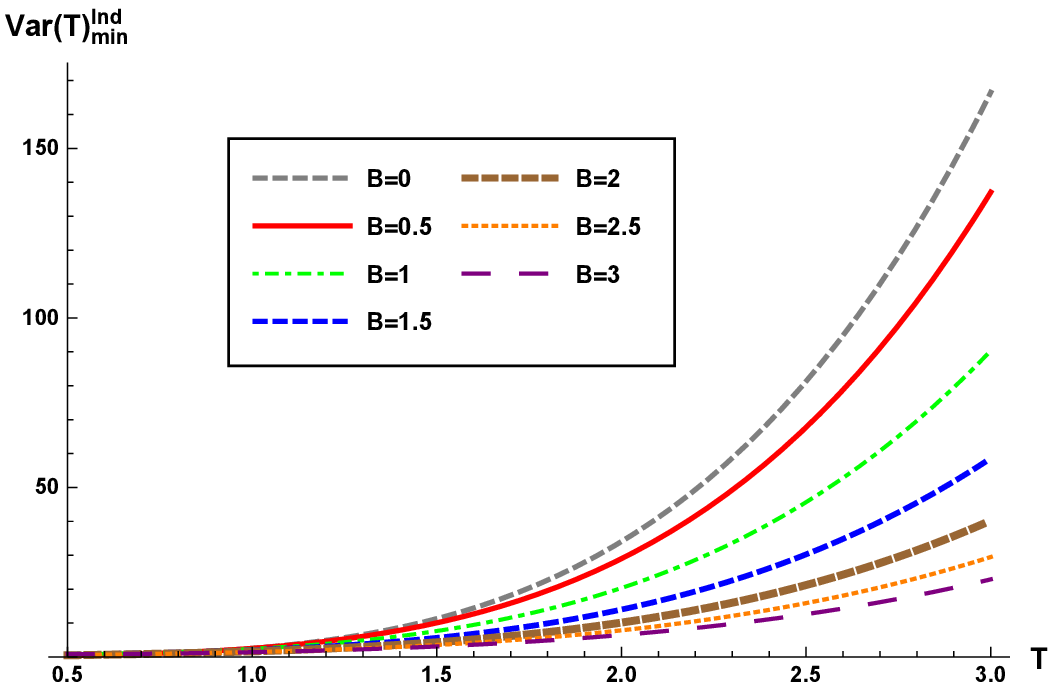}
    \end{minipage}

    \caption{\sf The minimal variances of individual estimates of parameters $B$ and $T$ with $J=1$.}
    \label{fig5}
\end{figure}
The results plotted in Fig.(\ref{fig5}) represent the evolution of the minimal values of the variance in the protocol of individual estimations the parameters $B$ and $T$. The behavior of these minimal variance are almost similar the results obtained by employing the simultaneous estimation strategy shown in Fig.(\ref{fig4}), but it presents an uncertainty of error in the precision of the optimal values of the parameters $B$ and $T$. This uncertainty can be quantified by the ratio between the minimal variance in individual estimation scenario and the minimal variance obtained in the simultaneous case. Using the equations (\ref{varminindB}) and (\ref{varminindT2}), it is easy to see that the equation (\ref{Raport}) gives

{\small \begin{equation}
\Gamma  = \frac{{\left( {2{e^{\beta B}} + \left( {1 + {e^{2\beta B}}} \right)\cosh \left( {\beta J} \right)} \right)\left( {\left( {1 + {e^{2\beta B}}} \right)\left( {{B^2} + {J^2}} \right)\cosh \left( {\beta J} \right) + 2\left( {\left( {{B^2} + {J^2}} \right){e^{\beta B}} - \left( {{e^{2\beta B}} - 1} \right)BJ\sinh \left( {\beta J} \right)} \right)} \right)}}{{2{J^2}{{\left( {1 + {e^{2\beta B}} + 2{e^{\beta B}}\cosh \left( {\beta J} \right)} \right)}^2}}}. \label{rapport2}
\end{equation}}
\begin{figure}[H]
    \begin{centering}
        \includegraphics{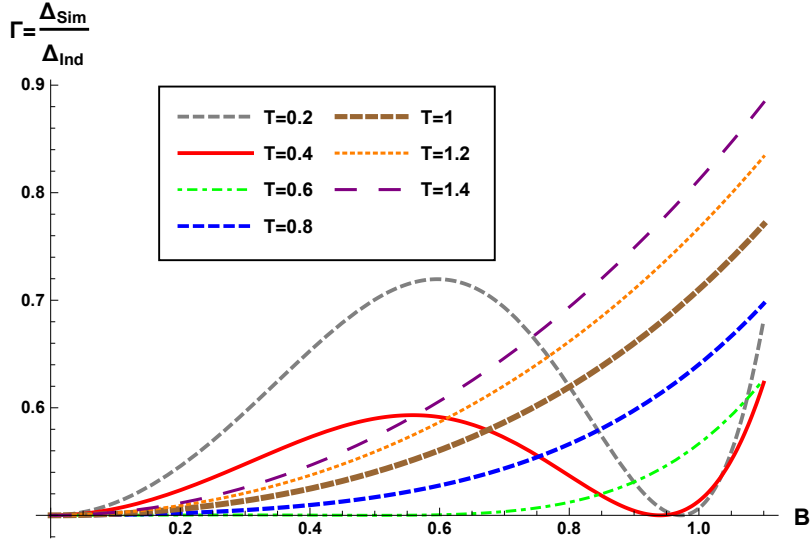}
        \caption{\sf The ratio between the minimal total variance of estimating the parameters $B$ and $T$ with $J=1$.}
        \label{Fig6}
    \end{centering}
\end{figure}
In Fig.(\ref{Fig6}), we plot the ratio $\Gamma$ (\ref{rapport2}) in the case where the coupling parameter $J=1$. As it can be seen from this figure, the minimal total variance corresponding to the simultaneous strategy is always less than the minimal total variance of the individual strategy, i.e., $\Delta _{\rm Sim} \leq \Delta _{\rm Ind}$. This confirms that the simultaneous estimation of the parameters $B$ and $T$ in the isotropic $XY$ model with a magnetic field can provide better precision than the individual estimation.

\section{Concluding remarks}
Quantum Fisher information matrix plays an essential role in extracting the maximum amount of information in order to get the best precision in measuring several physical quantities. Thus making it possible to find the optimal states of the system which correspond to optimal values of the estimated parameters. In this work, we have studied multiparametric estimation strategy in quantum metrology by focusing on two variants of the Heisenberg $XY$ model. The first one concerns the anisotropic $XY$ model and the second scenario deals with isotropic $XY$ model embedded in a magnetic field. We find the multiparameter quantum Cramér–Rao bound for simultaneous and individual estimation of the temperature, anisotropic parameter and magnetic field using the concept of quantum Fisher information matrix. In addition, we have compared simultaneous and individual estimation strategies. We have found that best precisions are obtained by employing the simultaneous estimation strategy.\par
The fact that the simultaneous estimation of several parameters in quantum metrology raises important questions. Indeed, it is natural to ask about the relation between the estimation precisions and the quantum correlations in enhancing the performance of a metrological protocol like for single-parameter estimation. Furthermore, as prolongation of this work, it will be interesting to investigate the dynamics of nonclassical correlations \cite{Slaoui2018,Slaoui22018,Kim2018}. In other words, it is interesting to study the characterization of quantum correlations in terms local quantum Fisher information and local quantum uncertainty and to study if they can provide the appropriate tools to examine the role of quantum correlations in multiparametric quantum metrology. In this paper, we focused only on the two-qubit systems. The analysis can be extended to multiqubit case. In the general case to compute the quantum Fisher information matrix one has to determine the inverse of the matrix $\Lambda  = {{\rho ^T} \otimes
    \mathbb{I} + \mathbb{I} \otimes \rho }$. The analytical expressions can be obtained using the Cholesky decomposition \cite{Krishnamoorthy2013}. We hope to report on this subject in a forthcoming work.

\end{document}